\documentclass[sigplan, 10pt, screen]{acmart}

\newif\ifisdraft
\isdraftfalse

\usepackage{xspace}
\usepackage[T1]{fontenc}
\usepackage[utf8]{inputenc}
\usepackage{xcolor}
\usepackage{listings}
\usepackage[normalem]{ulem}
\usepackage{amsfonts}

\usepackage{caption}
\usepackage{subcaption}

\usepackage[linesnumbered,ruled,vlined]{algorithm2e}

\definecolor{listing-comment}{rgb}{0.68,0.68,0.68}
\definecolor{listing-string}{rgb}{0, 0, 0.5}%
\definecolor{listing-keyword-traits}{rgb}{0.75, 0, 0}
\definecolor{listing-keyword-types}{rgb}{0, 0.5, 0}
\definecolor{listing-keyword-const}{rgb}{0, 0.5, 0}
\definecolor{listing-keyword-macro}{rgb}{0, 0, 0.75}
\definecolor{listing-keyword}{rgb}{0, 0, 0}
\definecolor{listing-identifier}{rgb}{0, 0, 0.75}

\newcommand{\lstbasicstyle}{\ttfamily\footnotesize}

\newcommand{\lstnumberstyle}{\tiny\color{gray}}

\newcommand{\lstkeywordstyle}{\bfseries\color{listing-keyword}}

\lstdefinestyle{common}
{
  firstnumber=1
  , basicstyle=\lstbasicstyle
  , identifierstyle=\color{listing-identifier}
  , commentstyle=\itshape\color{listing-comment}
  , stringstyle=\color{listing-string}
  , keywordstyle=\lstkeywordstyle
  , extendedchars=false
  , tabsize=2
  , showtabs=false
  , showspaces=false
  , showstringspaces=false
  , numbers=left
  , stepnumber=2
  , numberfirstline=false,
  , numberstyle=\lstnumberstyle
  , numbersep=5pt
  , captionpos=b
  , breaklines=true
  , breakatwhitespace=true
  , breakautoindent=true
  , frame=none
  , rulecolor=\color{black}
  , columns=[c]fixed
  , keepspaces=true
 , escapeinside={(*@}{@*)}
}

\lstdefinestyle{coloredProlog}
{
  language=Prolog
  , style = common
  , morecomment=[l]{\//},
}

\lstnewenvironment{codeprolog}
  {\lstset{style=coloredProlog}}
  {}

\lstnewenvironment{codeprologfloat}
  {\lstset{style=coloredProlog,float=ht}}
  {}

\lstdefinestyle{coloredC}
{
  language=C
  , style = common
  , morekeywords={uint8_t, uint16_t, uint32_t, uint64_t, inline}
  , morecomment=[l]{\//},
}

\lstnewenvironment{codec}
  {\lstset{style=coloredC}}
  {}

\lstnewenvironment{codecfloat}
  {\lstset{style=coloredC,float=ht}}
  {}

\lstdefinelanguage{Rust}{%
    sensitive%
  , morecomment=[l]{//}%
  , morecomment=[s]{/*}{*/}%
  , moredelim=[s][{\itshape\color[rgb]{0,0,0.75}}]{\#[}{]}%
  , morestring=[b]{"}%
  , alsodigit={}%
  , alsoother={}%
  , alsoletter={!}%
  , morekeywords={break, continue, else, for, if, in, loop, match, return, while}  
  , morekeywords={as, const, let, move, mut, ref, static}  
  , morekeywords={dyn, enum, fn, impl, Self, self, struct, trait, type, union, use, where}  
  , morekeywords={crate, extern, mod, pub, super}  
  , morekeywords={unsafe}  
  , morekeywords={abstract, alignof, become, box, do, final, macro, offsetof, override, priv, proc, pure, sizeof, typeof, unsized, virtual, yield}  
  , morekeywords=[2]{Add, AddAssign, Any, AsciiExt, AsInner, AsInnerMut, AsMut, AsRawFd, AsRawHandle, AsRawSocket, AsRef, Binary, BitAnd, BitAndAssign, Bitor, BitOr, BitOrAssign, BitXor, BitXorAssign, Borrow, BorrowMut, Boxed, BoxPlace, BufRead, BuildHasher, CastInto, CharExt, Clone, CoerceUnsized, CommandExt, Copy, Debug, DecodableFloat, Default, Deref, DerefMut, DirBuilderExt, DirEntryExt, Display, Div, DivAssign, DoubleEndedIterator, DoubleEndedSearcher, Drop, EnvKey, Eq, Error, ExactSizeIterator, ExitStatusExt, Extend, FileExt, FileTypeExt, Float, Fn, FnBox, FnMut, FnOnce, Freeze, From, FromInner, FromIterator, FromRawFd, FromRawHandle, FromRawSocket, FromStr, FullOps, FusedIterator, Generator, Hash, Hasher, Index, IndexMut, InPlace, Int, Into, IntoCow, IntoInner, IntoIterator, IntoRawFd, IntoRawHandle, IntoRawSocket, IsMinusOne, IsZero, Iterator, JoinHandleExt, LargeInt, LowerExp, LowerHex, MetadataExt, Mul, MulAssign, Neg, Not, Octal, OpenOptionsExt, Ord, OsStrExt, OsStringExt, Packet, PartialEq, PartialOrd, Pattern, PermissionsExt, Place, Placer, Pointer, Product, Put, RangeArgument, RawFloat, Read, Rem, RemAssign, Seek, Shl, ShlAssign, Shr, ShrAssign, Sized, SliceConcatExt, SliceExt, SliceIndex, Stats, Step, StrExt, Sub, SubAssign, Sum, Sync, TDynBenchFn, Terminal, Termination, ToOwned, ToSocketAddrs, ToString, Try, TryFrom, TryInto, UnicodeStr, Unsize, UpperExp, UpperHex, WideInt, Write}
  , morekeywords=[2]{Send}  
  , morekeywords=[3]{bool, char, f32, f64, i8, i16, i32, i64, isize, str, u8, u16, u32, u64, unit, usize, i128, u128}  
  , morekeywords=[4]{Err, false, None, Ok, Some, true}  
  , morekeywords=[3]{AccessError, Adddf3, AddI128, AddoI128, AddoU128, ADDRESS, ADDRESS64, addrinfo, ADDRINFOA, AddrParseError, Addsf3, AddU128, advice, aiocb, Alignment, AllocErr, AnonPipe, Answer, Arc, Args, ArgsInnerDebug, ArgsOs, Argument, Arguments, ArgumentV1, Ashldi3, Ashlti3, Ashrdi3, Ashrti3, AssertParamIsClone, AssertParamIsCopy, AssertParamIsEq, AssertUnwindSafe, AtomicBool, AtomicPtr, Attr, auxtype, auxv, BackPlace, BacktraceContext, Barrier, BarrierWaitResult, Bencher, BenchMode, BenchSamples, BinaryHeap, BinaryHeapPlace, blkcnt, blkcnt64, blksize, BOOL, boolean, BOOLEAN, BoolTrie, BorrowError, BorrowMutError, Bound, Box, bpf, BTreeMap, BTreeSet, Bucket, BucketState, Buf, BufReader, BufWriter, Builder, BuildHasherDefault, BY, BYTE, Bytes, CannotReallocInPlace, cc, Cell, Chain, CHAR, CharIndices, CharPredicateSearcher, Chars, CharSearcher, CharsError, CharSliceSearcher, CharTryFromError, Child, ChildPipes, ChildStderr, ChildStdin, ChildStdio, ChildStdout, Chunks, ChunksMut, ciovec, clock, clockid, Cloned, cmsgcred, cmsghdr, CodePoint, Color, ColorConfig, Command, CommandEnv, Component, Components, CONDITION, condvar, Condvar, CONSOLE, CONTEXT, Count, Cow, cpu, CRITICAL, CStr, CString, CStringArray, Cursor, Cycle, CycleIter, daddr, DebugList, DebugMap, DebugSet, DebugStruct, DebugTuple, Decimal, Decoded, DecodeUtf16, DecodeUtf16Error, DecodeUtf8, DefaultEnvKey, DefaultHasher, dev, device, Difference, Digit32, DIR, DirBuilder, dircookie, dirent, dirent64, DirEntry, Discriminant, DISPATCHER, Display, Divdf3, Divdi3, Divmoddi4, Divmodsi4, Divsf3, Divsi3, Divti3, dl, Dl, Dlmalloc, Dns, DnsAnswer, DnsQuery, dqblk, Drain, DrainFilter, Dtor, Duration, DwarfReader, DWORD, DWORDLONG, DynamicLibrary, Edge, EHAction, EHContext, Elf32, Elf64, Empty, EmptyBucket, EncodeUtf16, EncodeWide, Entry, EntryPlace, Enumerate, Env, epoll, errno, Error, ErrorKind, EscapeDebug, EscapeDefault, EscapeUnicode, event, Event, eventrwflags, eventtype, ExactChunks, ExactChunksMut, EXCEPTION, Excess, ExchangeHeapSingleton, exit, exitcode, ExitStatus, Failure, fd, fdflags, fdsflags, fdstat, ff, fflags, File, FILE, FileAttr, filedelta, FileDesc, FilePermissions, filesize, filestat, FILETIME, filetype, FileType, Filter, FilterMap, Fixdfdi, Fixdfsi, Fixdfti, Fixsfdi, Fixsfsi, Fixsfti, Fixunsdfdi, Fixunsdfsi, Fixunsdfti, Fixunssfdi, Fixunssfsi, Fixunssfti, Flag, FlatMap, Floatdidf, FLOATING, Floatsidf, Floatsisf, Floattidf, Floattisf, Floatundidf, Floatunsidf, Floatunsisf, Floatuntidf, Floatuntisf, flock, ForceResult, FormatSpec, Formatted, Formatter, Fp, FpCategory, fpos, fpos64, fpreg, fpregset, FPUControlWord, Frame, FromBytesWithNulError, FromUtf16Error, FromUtf8Error, FrontPlace, fsblkcnt, fsfilcnt, fsflags, fsid, fstore, fsword, FullBucket, FullBucketMut, FullDecoded, Fuse, GapThenFull, GeneratorState, gid, glob, glob64, GlobalDlmalloc, greg, group, GROUP, Guard, GUID, Handle, HANDLE, Handler, HashMap, HashSet, Heap, HINSTANCE, HMODULE, hostent, HRESULT, id, idtype, if, ifaddrs, IMAGEHLP, Immut, in, in6, Incoming, Infallible, Initializer, ino, ino64, inode, input, InsertResult, Inspect, Instant, int16, int32, int64, int8, integer, IntermediateBox, Internal, Intersection, intmax, IntoInnerError, IntoIter, IntoStringError, intptr, InvalidSequence, iovec, ip, IpAddr, ipc, Ipv4Addr, ipv6, Ipv6Addr, Ipv6MulticastScope, Iter, IterMut, itimerspec, itimerval, jail, JoinHandle, JoinPathsError, KDHELP64, kevent, kevent64, key, Key, Keys, KV, l4, LARGE, lastlog, launchpad, Layout, Lazy, lconv, Leaf, LeafOrInternal, Lines, LinesAny, LineWriter, linger, linkcount, LinkedList, load, locale, LocalKey, LocalKeyState, Location, lock, LockResult, loff, LONG, lookup, lookupflags, LookupHost, LPBOOL, LPBY, LPBYTE, LPCSTR, LPCVOID, LPCWSTR, LPDWORD, LPFILETIME, LPHANDLE, LPOVERLAPPED, LPPROCESS, LPPROGRESS, LPSECURITY, LPSTARTUPINFO, LPSTR, LPVOID, LPWCH, LPWIN32, LPWSADATA, LPWSAPROTOCOL, LPWSTR, Lshrdi3, Lshrti3, lwpid, M128A, mach, major, Map, mcontext, Metadata, Metric, MetricMap, mflags, minor, mmsghdr, Moddi3, mode, Modsi3, Modti3, MonitorMsg, MOUNT, mprot, mq, mqd, msflags, msghdr, msginfo, msglen, msgqnum, msqid, Muldf3, Mulodi4, Mulosi4, Muloti4, Mulsf3, Multi3, Mut, Mutex, MutexGuard, MyCollection, n16, NamePadding, NativeLibBoilerplate, nfds, nl, nlink, NodeRef, NoneError, NonNull, NonZero, nthreads, NulError, OccupiedEntry, off, off64, oflags, Once, OnceState, OpenOptions, Option, Options, OptRes, Ordering, OsStr, OsString, Output, OVERLAPPED, Owned, Packet, PanicInfo, Param, ParseBoolError, ParseCharError, ParseError, ParseFloatError, ParseIntError, ParseResult, Part, passwd, Path, PathBuf, PCONDITION, PCONSOLE, Peekable, PeekMut, Permissions, PhantomData, pid, Pipes, PlaceBack, PlaceFront, PLARGE, PoisonError, pollfd, PopResult, port, Position, Powidf2, Powisf2, Prefix, PrefixComponent, PrintFormat, proc, Process, PROCESS, processentry, protoent, PSRWLOCK, pthread, ptr, ptrdiff, PVECTORED, Queue, radvisory, RandomState, Range, RangeFrom, RangeFull, RangeInclusive, RangeMut, RangeTo, RangeToInclusive, RawBucket, RawFd, RawHandle, RawPthread, RawSocket, RawTable, RawVec, Rc, ReadDir, Receiver, recv, RecvError, RecvTimeoutError, ReentrantMutex, ReentrantMutexGuard, Ref, RefCell, RefMut, REPARSE, Repeat, Result, Rev, Reverse, riflags, rights, rlim, rlim64, rlimit, rlimit64, roflags, Root, RSplit, RSplitMut, RSplitN, RSplitNMut, RUNTIME, rusage, RwLock, RWLock, RwLockReadGuard, RwLockWriteGuard, sa, SafeHash, Scan, sched, scope, sdflags, SearchResult, SearchStep, SECURITY, SeekFrom, segment, Select, SelectionResult, sem, sembuf, send, Sender, SendError, servent, sf, Shared, shmatt, shmid, ShortReader, ShouldPanic, Shutdown, siflags, sigaction, SigAction, sigevent, sighandler, siginfo, Sign, signal, signalfd, SignalToken, sigset, sigval, Sink, SipHasher, SipHasher13, SipHasher24, size, SIZE, Skip, SkipWhile, Slice, SmallBoolTrie, sockaddr, SOCKADDR, sockcred, Socket, SOCKET, SocketAddr, SocketAddrV4, SocketAddrV6, socklen, speed, Splice, Split, SplitMut, SplitN, SplitNMut, SplitPaths, SplitWhitespace, spwd, SRWLOCK, ssize, stack, STACKFRAME64, StartResult, STARTUPINFO, stat, Stat, stat64, statfs, statfs64, StaticKey, statvfs, StatVfs, statvfs64, Stderr, StderrLock, StderrTerminal, Stdin, StdinLock, Stdio, StdioPipes, Stdout, StdoutLock, StdoutTerminal, StepBy, String, StripPrefixError, StrSearcher, subclockflags, Subdf3, SubI128, SuboI128, SuboU128, subrwflags, subscription, Subsf3, SubU128, Summary, suseconds, SYMBOL, SYMBOLIC, SymmetricDifference, SyncSender, sysinfo, System, SystemTime, SystemTimeError, Take, TakeWhile, tcb, tcflag, TcpListener, TcpStream, TempDir, TermInfo, TerminfoTerminal, termios, termios2, TestDesc, TestDescAndFn, TestEvent, TestFn, TestName, TestOpts, TestResult, Thread, threadattr, threadentry, ThreadId, tid, time, time64, timespec, TimeSpec, timestamp, timeval, timeval32, timezone, tm, tms, ToLowercase, ToUppercase, TraitObject, TryFromIntError, TryFromSliceError, TryIter, TryLockError, TryLockResult, TryRecvError, TrySendError, TypeId, U64x2, ucontext, ucred, Udivdi3, Udivmoddi4, Udivmodsi4, Udivmodti4, Udivsi3, Udivti3, UdpSocket, uid, UINT, uint16, uint32, uint64, uint8, uintmax, uintptr, ulflags, ULONG, ULONGLONG, Umoddi3, Umodsi3, Umodti3, UnicodeVersion, Union, Unique, UnixDatagram, UnixListener, UnixStream, Unpacked, UnsafeCell, UNWIND, UpgradeResult, useconds, user, userdata, USHORT, Utf16Encoder, Utf8Error, Utf8Lossy, Utf8LossyChunk, Utf8LossyChunksIter, utimbuf, utmp, utmpx, utsname, uuid, VacantEntry, Values, ValuesMut, VarError, Variables, Vars, VarsOs, Vec, VecDeque, vm, Void, WaitTimeoutResult, WaitToken, wchar, WCHAR, Weak, whence, WIN32, WinConsole, Windows, WindowsEnvKey, winsize, WORD, Wrapping, wrlen, WSADATA, WSAPROTOCOL, WSAPROTOCOLCHAIN, Wtf8, Wtf8Buf, Wtf8CodePoints, xsw, xucred, Zip, zx}
  , morekeywords=[5]{assert!, assert_eq!, assert_ne!, cfg!, column!, compile_error!, concat!, concat_idents!, debug_assert!, debug_assert_eq!, debug_assert_ne!, env!, eprint!, eprintln!, file!, format!, format_args!, include!, include_bytes!, include_str!, line!, module_path!, option_env!, panic!, print!, println!, select!, stringify!, thread_local!, try!, unimplemented!, unreachable!, vec!, write!, writeln!}  
}

\lstdefinestyle{coloredRust}
  {
    language=Rust
    , style = common
    , keywordstyle=\bfseries
    , keywordstyle=[2]\color[rgb]{0.75, 0, 0}
    , keywordstyle=[3]\color[rgb]{0, 0.5, 0}
    , keywordstyle=[4]\color[rgb]{0, 0.5, 0}
    , keywordstyle=[5]\color[rgb]{0, 0, 0.75}
    , morecomment=[l]{\//},
  }

\lstnewenvironment{coderust}
  {\lstset{style=coloredRust}}
  {}

\lstnewenvironment{coderustfloat}
  {\lstset{style=coloredRust,float=ht}}
  {}

\lstdefinelanguage{Vrs}{%
  sensitive%
, morecomment=[l]{//}%
, morecomment=[s]{/*}{*/}%
, moredelim=[s][{\itshape\color[rgb]{0,0,0.75}}]{\#[}{]}%
, morestring=[b]{"}%
, alsodigit={}%
, alsoother={}%
, alsoletter={!}%
, morekeywords={break, continue, else, for, if, in, loop, match, return, while}  
, morekeywords={as, const, let, move, mut, ref, static}  
, morekeywords={unit, fn, requires, ensures, forall, exists}  
, morekeywords={Register, Memory, MMIO, CPURegister}  
, morekeywords=[2]{Layout, ReadAction, WriteAction}
, morekeywords=[3]{inaddr, outaddr, flags,size, int, nat}  
, morekeywords=[4]{true, false}  
, morekeywords=[4]{state, interface}  
}

\lstdefinestyle{coloredvrs}
{
  language=Vrs
  , style = common
  , keywordstyle=\bfseries
  , keywordstyle=[2]\color[rgb]{0.75, 0, 0}
  , keywordstyle=[3]\color[rgb]{0, 0.5, 0}
  , keywordstyle=[4]\color[rgb]{0, 0.5, 0}
  , keywordstyle=[5]\color[rgb]{0, 0, 0.75}
  , morecomment=[l]{\//},
}

\lstnewenvironment{codevrs}
  {\lstset{style=coloredvrs}}
  {}

\lstnewenvironment{codevrsfloat}[1]
  {\lstset{style=coloredvrs,float=ht, #1}}
  {}

\lstdefinelanguage{Scheme}{%
  morekeywords=[1]{define, define-syntax, define-macro, lambda, define-stream, stream-lambda},
  morekeywords=[2]{begin, call-with-current-continuation, call/cc,
    call-with-input-file, call-with-output-file, case, cond,
    do, else, for-each, if,
    let*, let, let-syntax, letrec, letrec-syntax,
    let-values, let*-values,
    and, or, not, delay, force,
    quasiquote, quote, unquote, unquote-splicing,
    map, fold, syntax, syntax-rules, eval, environment, query },
  morekeywords=[3]{import, export},
  alsodigit=!\$\%&*+-./:<=>?@^_~,
  sensitive=true,
  morecomment=[l]{;},
  morecomment=[s]{\#|}{|\#},
  morestring=[b]",
  basicstyle=\small\ttfamily,
  keywordstyle=\bf\ttfamily\color[rgb]{0,.3,.7},
  commentstyle=\color[rgb]{0.133,0.545,0.133},
  stringstyle={\color[rgb]{0.75,0.49,0.07}},
  upquote=true,
  breaklines=true,
  breakatwhitespace=true,
  literate=*{`}{{`}}{1},
  showstringspaces=false
}

\lstdefinestyle{coloredscheme}
{
  language=Vrs
  , style = common
  , keywordstyle=\bfseries
  , keywordstyle=[2]\color[rgb]{0.75, 0, 0}
  , keywordstyle=[3]\color[rgb]{0, 0.5, 0}
  , keywordstyle=[4]\color[rgb]{0, 0.5, 0}
  , keywordstyle=[5]\color[rgb]{0, 0, 0.75}
  , morecomment=[l]{\//},
}

\lstnewenvironment{codescheme}
  {\lstset{style=coloredscheme}}
  {}

\usepackage{marginnote}

\ifisdraft

\newcommand{\nbc}[3]{
    {{\colorbox{#3}{\bfseries\sffamily\scriptsize\textcolor{white}{#1}}}}
    {\textcolor{#3}{\sf\small$\blacktriangleright~$\textit{#2}$~\blacktriangleleft$}}}

\else
\newcommand{\nbc}[3]{}
\fi

\newcommand{\system}{\emph{OSmosis}\xspace}

\newcommand{\newPD}{\emph{newPD}\xspace}
\newcommand{\newPDFC}{\emph{clonePD}\xspace}

\acmYear{}
\copyrightyear{}
\acmConference[]{}
\acmBooktitle{}
\acmPrice{}
\acmDOI{}
\acmISBN{}

\renewcommand\footnotetextcopyrightpermission[1]{} 
\begin{document}




\title{OSmosis: No more Déjà vu in OS isolation}
\acmSubmissionID{98}


\author{Sidhartha Agrawal}
\orcid{0000-0003-3194-6037}
\affiliation{%
\institution{University of British Columbia}
\city{Vancouver}
\country{Canada}
}

\definecolor{purplecolor}{rgb}{0.56,0.10,0.63}
\newcommand\SA[1]{\nbc{SA}{#1}{purplecolor}}

\author{Reto Achermann}
\orcid{0000-0003-3263-7236}
\affiliation{%
\institution{University of British Columbia}
\city{Vancouver}
\country{Canada}
}

\definecolor{greencolor}{rgb}{0.11,0.63,0.14}
\newcommand\RA[1]{\nbc{RA}{#1}{greencolor}}

\author{Margo Seltzer}
\orcid{0000-0002-2165-4658}
\affiliation{%
\institution{University of British Columbia}
\city{Vancouver}
\country{Canada}
}

\definecolor{bluecolor}{rgb}{0.11,0.32,1.00}
\newcommand\MIS[1]{\nbc{MIS}{#1}{bluecolor}}

\renewcommand{\shortauthors}{Agrawal, et al.}

\newcommand{\change}[2][]
	{{\color{orange}{#2}}}


\begin{abstract}

Operating systems provide an abstraction layer between the
hardware and higher-level software.
Many abstractions, such as threads, processes, containers,
and virtual machines, are mechanisms to provide isolation.
New application scenarios frequently introduce new isolation
mechanisms.
%
Implementing each isolation
mechanism as an independent abstraction makes it difficult
to reason about the state and resources shared among different
tasks, leading to security vulnerabilities and performance interference.

We present \system, an isolation model
that expresses the precise level of resource sharing,
a framework in which to implement isolation mechanisms based on the model,
and an implementation of the framework on seL4.
%
The \system model lets the user determine the degree of isolation guarantee that
they need from the system.
This determination empowers
developers to make informed decisions about isolation and
performance trade-offs,
and the framework enables them to create mechanisms
with the desired degree of isolation.
\end{abstract}


\maketitle


\section{Introduction}
\label{sec:introduction}

From the moment that more than one person wanted to use a computer at the same time
(some 60 years ago), the systems community has developed myriad of techniques to facilitate safe multiplexing.
The community continues to struggle with how to provide the right degree of sharing and isolation
for a given application and its users
~\cite{corbato1962experimental,meyer1970virtual,barham2003xen,jing2022orbitz,litton2016light,openVZ,namespacesLinux,jailsFreeBSD,zonesSolaris,cloneLinux,dockerSite}.
Even more problematic is that there is no clear understanding of the isolation levels provided by different mechanisms.
Perhaps more fundamentally, given an isolation mechanism, it is not immediately clear what the application state
consists of and thus, what parts of the application's state are shared with or isolated from other applications.
While some application state is known (e.g., heap, code, data, and less obvious the kernel),
there exists a significant amount of unknown
state that the application is inadvertently sharing
with other applications (e.g., system-level services (\autoref{sec:motivation-state})).
Worse, we lack a common vocabulary to describe an application's resources, including its
known and unknown software
state.
This has led to many problems ranging from performance anomalies due to unintentional sharing
and overheads from too much isolation to security vulnerabilities caused by unintentional
sharing of known and unknown software
state~\cite{dosCVE,YangSharedDemonInTheKernel}

We claim that there is a need for a principled way to talk about isolation and sharing,
and a framework upon which to build implementations.
Our hypothesis is
that all OS mechanisms can be described
as a set of \textbf{resources} and the \textbf{relationships} describing
dependencies among them.
Resources can be the virtual memory an application uses,
the files to which it has access, the OS state it can query,
etc.
Resources lie on a sharing spectrum ranging from wholly shared to completely isolated.
The metric that defines this spectrum is the distance to the first
common resource found in the resource relationships of two entities
(i.e., processes, containers, virtual machines).
For example, two threads are on the shared end of the spectrum,
because they share a virtual address space resource.
Two processes running on two different virtual machines are more isolated,
because the first resource they share is state maintained by a hypervisor.

We present \system, which is composed of two parts.
First, the \system model (\autoref{sec:model::model}) describes the types of entities in a
system and how their sharing lies on a spectrum (\autoref{sec:model::spectrum}).
The model gives us a principled way to express isolation and sharing.
Second, the \system framework (\autoref{sec:framework}) describes the
required OS mechanisms and tools for implementing the isolation model.
The \system framework allows us to select a specific point in a high-dimensional space,
where the different axes correspond to the different resources (e.g., physical
memory, CPUs).

We have built a prototype system on the capabilities-based seL4 microkernel~\cite{seL4Website}
that implements parts of our framework.
We have built two existing and two new mechanisms with our framework.
In contrast to existing implementations,  we use the same set of
building blocks for every mechanism.

\system lets us model the levels of isolation for each subsystem independently.
For instance, if we are more concerned about attacks in the networking stack,
we can give the networking stack stronger isolation than
the file system stack.
When subsystems are tightly coupled,
(e.g., the virtual memory and file systems),
increasing the isolation level for one
raises the isolation level for the other,
but this tight coupling does not exist between all subsystems.

With \system, we can model the isolation requirements of a given
application and then easily build it using the framework.
This is especially exciting with the rise of serverless architectures, where
the simple choice between VM and container has become significantly more complicated,
and myriad new container/VM hybrids emerge regularly~\cite{kataContainers,unikraft,whitaker2002denali,shillaker2020faasm,sartakov2022capvm}.
There is no one-size-fits-all, and for a given application,
one might want to pick different degrees of isolation/sharing
between applications running on the same machine.

\section{Motivation}
\label{sec:motivation}

New mechanisms are often motivated by
one of: the emergence of a new use case,
improving the performance of an existing use case,
or defending against a security vulnerability.
However, the solutions always use isolation as a tool.
For example, they reserve resources (memory, storage, CPU time)~\cite{cgroupsLinux},
restrict access to unneeded state (kernel)~\cite{shillaker2020faasm}, or
share underlying state (drivers)~\cite{unikraft,agache2020firecracker} to improve performance.
Similarly, they increase the isolation of resources or underlying state
to build defenses.
Given the importance of varying isolation, it is useful to have a clear understanding of
which resources and state are shared among applications.

\subsection{New use cases}
The systems community has developed many different isolation mechanisms in the last decade
\cite{hsu2016smv,Jing2022Orbit,chen2016shreds,litton2016light}, and
each provides a slightly different degree of isolation for a different resource,
such as virtual memory, open files, performance, etc.
Shreds~\cite{chen2016shreds},
Secure Memory Views (SMV)~\cite{hsu2016smv},
and Light Weight Contexts (LwC)~\cite{litton2016light}
focus on providing compartmentalization within an address space,
while LwCs also provide compartmentalization of some kernel state (e.g., file descriptor table)
within the same process.

Whenever a new scenario arises, a solution is built to fit it, but there is no principled way
to describe and implement these solutions, making it difficult to formally distinguish different
solutions from one another.
With new paradigms such as Function-as-a-Service, we see many more
mechanisms arise \cite{razaSoKFunctionAsAServiceApplication2021,shillaker2020faasm}.
Some restrict access to underlying state ~\cite{shillaker2020faasm}, not needed by the function.
Others run multiple functions in the same process and use additional
mechanisms to create intra-address-space isolation~\cite{sartakov2022capvm}.
Although these mechanisms provide incremental isolation levels, their implementations are not incremental.
A new implementation is prone to bugs since it cannot take advantage of years of testing on existing mechanisms
\cite{lxcCVE,kvmCVE,chrootCWE}.

Furthermore, some organizations do not have the engineering resources to develop a new mechanism from scratch,
so applications are retrofitted into existing mechanisms.
Application developers might use a mechanism that provides weaker
isolation than desired, leaving them vulnerable to exploits.
Alternatively, they might use a mechanism with
overly strong isolation and pay more for their deployment in a shared cloud environment.

\subsection{Unintentional resource sharing}
\label{sec:motivation-state}
The lack of clarity about the extent of sharing between two applications
is also a source of security vulnerabilities.
Even when applications appear isolated, such as in the case of a container,
they still share kernel state.
Since namespaces do not isolate all the visible state in the kernel,
some state still leaks across container boundaries
(e.g., the open file table) leading to denial-of-service attacks on other
applications on the same host~\cite{YangSharedDemonInTheKernel}.
Additionally, since the container infrastructure and the kernel run in the same address space,
a simple buffer overflow in one part of the kernel can bring down the shared kernel
and both containers.
Lightweight VMs such as FireCracker~\cite{agache2020firecracker}
and KataContainers~\cite{kataContainers}
are more secure alternatives to containers,
providing the security of VMs with the overhead of containers.
However, they achieve this performance by having the host OS
provide functionality (e.g., drivers) for all VMs instead of the guest OS.
Unfortunately, this leads to more shared state in the host kernel.
Just as a shared kernel exposed issues with state leakage (for containers),
shared drivers in the host kernel can do the same (for VMs).


\section{\system Isolation Model}
\label{sec:model}

We now present the \system isolation model, the types of queries possible
on the model, and how they lead to a precise definition of the isolation spectrum.

\subsection{Model}
\label{sec:model::model}

\autoref{lst:model-formal} shows the \system model.
It consists of a \emph{system} and a \emph{resource relation}.
A system consists of a set of \emph{protection domains} and a set of \emph{resources}.
Protection domains (PD) correspond to active entities (e.g., processes, threads, virtual machines).
A PD has a set of resources and a \emph{resource directory}.
Resources are passive entities and can be either physical (e.g., RAM, CPUs or devices)
or virtual (e.g., virtual memory region, file, socket).
Physical resources all derive from tangible elements; virtual resources can
be created by a PD.
Both types of resources can be partitioned into smaller resources.
The resource directory is a dictionary, keyed by a resource, that identifies
the PD responsible for satisfying a request for a resource that the current
PD does not possess. For example, a user-level process wanting to allocate some
memory will call \texttt{mmap()} to request more virtual memory resources. This
corresponds to a lookup in the PD's resource directory for virtual memory resources
and then requesting more virtual memory from the PD to which that resource maps.

\lstset{basicstyle=\footnotesize\ttfamily}
\begin{lstlisting}[
    float=h,
    caption={\system Isolation Model},
    captionpos=b,
    escapeinside={(*}{*)},
    label={lst:model-formal}]
(*\textbf{System}*) = { pds:Set<PD>, res:Set<Resource> }
(*\textbf{Resource Relation}*) :: Resource x Resource
(*\textbf{PD}*) = { res:Set<Resource>, rdir:ResourceDirectory }
(*\textbf{Resource}*) = Virtual Resource | Physical Resource
(*\textbf{Virtual Resource}*) = Virtual Memory, File, ...
(*\textbf{Physical Resource}*) = RAM, Blocks, NIC, ...
(*\textbf{ResourceDirectory<Resource>}*) = PD for a resource
\end{lstlisting}

The \emph{resource relation} describes dependencies between two resources.
Each traversal of the relation is called a hop.
There are three types of resource relations.
The first is due to the system topology and does not change.
For example, the contents of DRAM may be loaded in the processor caches or sent over the memory bus.
The second type is added by system software.
For example, the page table keeps track of which virtual memory pages
are mapped to a physical page.
And finally, a resource depends on the underlying resource from which
it was allocated.
For example, a virtual page depends on the virtual
address space (resource) from which it was allocated.

\begin{figure}[t]
    \centering
    \includegraphics[width=1.0\columnwidth]{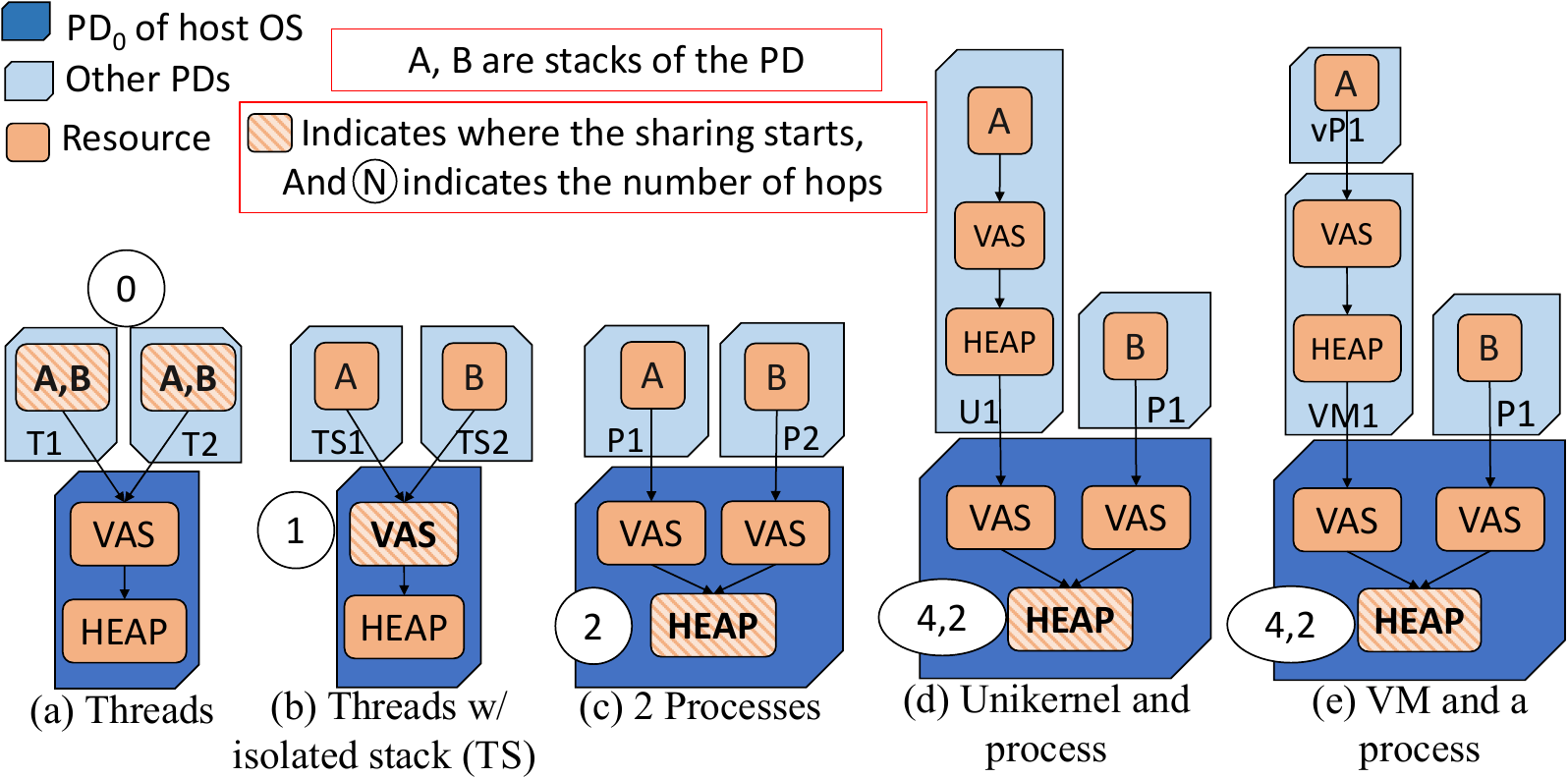}
    \caption{Five mechanisms modeled with \system
    }
    \label{fig:model}
\end{figure}

We show the flexibility of our model by describing five scenarios
in \autoref{fig:model}:
1) two threads in a process,
2) two threads with isolated stacks,
3) two processes,
4) a unikernel and a process,
5) a virtual machine and a process -- all running on the same monolithic OS (e.g., Linux).
In this example, we focus only on memory resources, but the concepts apply to
all types of resources.
The patterned boxes indicate where the sharing begins,
and the ovals indicate the number of hops at which the sharing happens.
Resources A and B represent the stack resource.
Two threads both have each other's stacks
in their protection domain (~\autoref{fig:model} (a)).
In 'threads with isolated stacks' (\autoref{fig:model} (b)),
each thread has access only to its own stack.
However, they are still allocated from the same address space.
Two processes (\autoref{fig:model} (c)) have separate address spaces,
but their virtual address space (VAS) data structures
in $PD_0$ depend on the kernel heap.
In the case of a unikernel (\autoref{fig:model} (d)), although there are
additional levels of abstraction,
address space management and the application are in the same PD.
In the case of a guestOS (i.e., virtual machine), a process
running on the VM is in a separate PD (\autoref{fig:model} (e)).

None of the PDs for threads, threads with isolated stacks, or processes
have direct access to physical resources.
Instead, when they need physical memory (i.e., on a page fault),
the Resource Directory indicates that $PD_0$ (i.e., the operating system)
will handle requests for physical memory.
In contrast, the guest OS handles such requests from
the process running inside the virtual machine,
while the host OS handles requests
from the hypervisor and its native process (P1).
When $PD_0$ maps a physical page to a virtual page, conceptually,
it adds an entry to the resource relation, even though in implementation,
this information is recorded in a page table.

\subsection{Queries}
\label{sec:model::queries}

\setlength{\abovedisplayshortskip}{4pt}
\setlength{\belowdisplayshortskip}{4pt}
\setlength{\abovedisplayskip}{5pt}
\setlength{\belowdisplayskip}{5pt}

%
We now define queries on the model to extract information about PDs,
their resources, and most importantly the relationship among resources
in different PDs.
In the next section, we show how to use this information to define the \emph{isolation spectrum}.

\textbf{NHopResources:}
The \textit{resource relation} lists the possible ``one-hop'' dependencies of a resource.
However, as we saw in the discussion of virtual addresses and processor caches,
there may be multiple levels of dependencies.
Thus, to identify all the resources on which a specific resource
depends, we compute the
\emph{n-hop transitive-reflexive} closure for n=INFINITY on the \textit{resource relation}.
\begin{align*}
    \textit{NHopResources} & :: \mathbb{N} \Rightarrow \textit{Set<Resource>}\Rightarrow  \textit{Set<Resource>}\\
    \textit{NHopResources} & :: n~R =  \bigcup_{r \in R } \textit{ResourceRelation}^n
\end{align*}

Referring back to~\autoref{fig:model}(c), consider the stack virtual memory region ($A$).
The stack's one-hop closure includes the VAS(in $PD_0$);
The VAS depends on the heap virtual memory resource of $PD_0$
from which its metadata was allocated;
its two-hop closure includes
cache sets (assuming virtually indexed caches) and the physical pages
that have been allocated to the virtual memory region.

\subsection{Isolation Spectrum}
\label{sec:model::spectrum}

Using the model and its queries, we can now define different forms of isolation as points on a \emph{spectrum}
and thus, we can quantify how isolated two PDs are from each other.\\

\noindent\textbf{NHopResourcesOfPD:}
We derive the n-hop resources of a PD by computing the NHopResource function
on the PD's resources unioned with the
($n-1$)-hop computation of NHopResourcesOfPD for each PD
in the resource directory.
\begin{align*}
    \textit{NHopResourcesOfPD}& :: \mathbb{N} \Rightarrow \textit{PD} \Rightarrow  \textit{Set<Resource>} \\
    \textit{NHopResourcesOfPD}& :: n~pd = (\textit{NHopResources}~n~pd.res)~\cup\\
                               \phantom{ } \bigcup_{p \in pd.rdir.values} & (\textit{NHopResourcesOfPD}~(n-1)~p)
\end{align*}
Note, this includes both the resources currently accessible as well as those
to which it may acquire access in the future. \\

\noindent\textbf{NHopShared:} We can now express the degree of sharing between two PDs by examining
the intersection of the sets produced by NHopResourcesOfPD for any values of n.
\begin{align*}
    \textit{NHopShared} ::&\mathbb{N} \Rightarrow \mathbb{N} \Rightarrow \textit{PD} \Rightarrow \textit{PD}\Rightarrow  \textit{Set<Resource>}\\
    \textit{NHopShared} ::&n_1~n_2~pd_1~pd_2 = \\
&\quad\quad\quad((\textit{NHopResourcesOfPD}~n_1~pd_1)~\cap \\
&\quad\quad\quad(\textit{NHopResourcesOfPD}~n_2~pd_2))
\end{align*}

\noindent\textbf{NHopIsolated:}
We say that two PDs are \emph{n-hop isolated} if they do not share any resources within $n$ hops of either PD, subject to an exclusion set, $\delta$.
The exclusion set $\delta$ is the set of resources the application does not care about.
When two processes do not care about sharing a cache or the file system, we add those resources to $\gamma$.
We say that a PD is \emph{NHopIsolated in the system}
if it is \emph{NHopIsolated} with every other PD.

\begin{align*}
    \textit{NHopIsolated} ::& \mathbb{N} \Rightarrow \mathbb{N} \Rightarrow\\
        &\textit{Set<Resource>} \Rightarrow \textit{PD} \Rightarrow \textit{PD}\Rightarrow bool\\
    \textit{NHopIsolated} ::& n_1~n_2~\delta~pd_1~pd_2 = \\
        &{NHopShared}~n_1~n_2~pd_1~pd_2 \subseteq \delta
\end{align*}

\noindent\textbf{IsolationLevel:}
Given two PDs, we define \emph{Isolation Level}
between them as the number of hops at which sharing begins.
We find the minimum value of $n_1$ or $n_2$ for which \emph{NHopIsolated} is false.
Taking the minimum of the tuple ensures proper accounting for asymmetric configurations.
In \autoref{fig:model}(e) the isolation level is 2 derived from $min(4, 2)$,
which means that sharing starts at 2 hops from at least one of the PDs.

\begin{align*}
    &\textit{IsolationLevel} :: \textit{PD} \Rightarrow \textit{PD} \Rightarrow \textit{Set<Resource>} \Rightarrow  \mathbb{N}\\
    &\textit{IsolationLevel} :: pd_1~pd_2~\delta = n~|~\forall n_1n_2 \\
    &\quad\quad\neg\textit{NHopIsolated}~n_1~n_2~\delta~pd_1~pd_2 \Longrightarrow~n~\leq min(n_1, n_2)
\end{align*}

One could argue that it's more useful to instead view isolation level
asymmetrically, i.e. from the perspective of each PD as opposed to between PDs.
Both perspectives have merit and we believe that more experience in implementing
various configurations will shed insight into which is more useful.
In either case, the model provides all the information necessary to engage in
this debate, and in fact, without the model, such debates cannot happen.
\section{\system Framework}
\label{sec:framework}

We now map the model described in the previous section to
the functionality required to realize it.

\subsection{Unified API}
\label{sec:framework_api}

Our model enables a unified API that can create any type of PD (inspired
by the \texttt{posix\_spawn} in Linux~\cite{posixSpawnLinux}).
\newPD takes a set of \emph{resources} and a \emph{resource directory}.
By default (i.e., if the resource directory is empty), a PD directs
requests for resources not given during its creation to the PD
that created it (e.g., processes redirect those requests to the OS).

\definecolor{mygreen}{rgb}{0,0.6,0}
\definecolor{mygray}{rgb}{0.5,0.5,0.5}
\definecolor{mymauve}{rgb}{0.58,0,0.82}
\lstset{
  basicstyle=\footnotesize\ttfamily,
  commentstyle=\color{mygreen},
  numberstyle=\tiny\color{mygray},
  numbers=right,
  numbersep=-10pt}

\begin{lstlisting}[
    float = b,
    caption={Process creation in \system},
    captionpos=b,
    label={lst:api-process},
    language=C]
 directory = currentResourceDirectory();
 vas = newVAS();

 // Get code, stack, heap, and vCPU resources
 code, heap, stack = load(vas, "binary");
 vCPU = newVCPU();
 resources = {code, heap, stack, vCPU};

 // Create the PD
 pdID = newPD(resources, directory);
\end{lstlisting}

\autoref{lst:api-process} shows the pseudo code to create a normal process in \system.
Since a process runs in a separate address space, we first create
a new virtual address space resource.
The \texttt{load} call reads a binary file and allocates from the \texttt{vas} resource
to create \texttt{code}, \texttt{heap}, and \texttt{stack} resources.

There are instances where the new PD is a close replica of an existing PD,
and the pseudo code in \autoref{lst:api-process} can be cumbersome.
Taking inspiration from \texttt{clone}~\cite{cloneLinux},
\newPDFC creates a new PD by calling an \emph{isolation function}
that defines how each resource and resource directory entry
is shared with its parent (or another PD) before calling \newPD.
It is not fundamental to our model or framework, but it is the
syntactic sugar that makes the model easier to use.
We have written a few \emph{isolation functions},
for example, one that creates a process or a thread with a
slightly different address space and one
where the resource directory entries for different types of resources
are different PDs.
We can also define \emph{isolation functions} that set up resources for the new PD
based on the degrees of isolation ($n_1, n_2$) for each resource and a $\delta$.
We envision having a suite of such template functions in a userspace library.

\subsection{Determining resource relations}
\label{sec:framework_resource_relation}
The resource relation (\autoref{sec:model::model}) makes it possible
to determine what underlying resources are shared.
The resource relation captures all dependencies between resources and thus can become quite large.
Yet, much of the information that is conceptually part of the resource relation
is already present in the system.
For example,
Linux provides the \texttt{sysfs}, \texttt{procfs}, \texttt{dev} file systems,
which describe system topology; page tables store
virtual address to physical address dependencies.
Dependencies between resources and the PDs that allocated them are
implied.
Designing an API to query the resource relation is straightforward;
building an efficient implementation of that query API is a more interesting
problem.
Fortunately, queries of the resource relation are not on the performance critical path
of normal operation.

\section{Implementation}
\label{sec:solution:impl}
We have implemented a small portion of \system -- dealing with memory resources --
using the capabilities-based microkernel seL4~\cite{seL4Website,sel4cp}.
We chose this microkernel as it has no existing abstractions for processes, containers, or virtual machines.
This lack of existing abstractions allows us to define the building blocks as we see fit.
\emph{Capabilities} and \emph{capability spaces}~\cite{seL4manual} map well to
\system{}'s resources and protection domains.
In seL4, the capability space is modeled as a tree, and, by sharing parts of the subtree with other
protection domain, we implement the sharing of resources amongst PDs.

\section{Discussion and Use cases}
\label{sec:disucussion_uc}

\system enables us to explore the space of isolation mechanisms in a principled way.
We discuss how the \emph{model} enables us to reason about isolation and the \emph{framework}
lets us build new abstractions quickly.

\subsection{Comparing Isolation level between PDs}
Viewing the systems as a collection of resources and relations
enables us to define queries on the model state
that can be used to precisely compare the level
of isolation between two PDs.
For instance, if we take the transitive closure of the resource relations starting at a PD,
we get a set of all the resources on which a PD depends.
Alternatively, we can restrict the number of relations to traverse (i.e., hops) to a small number
and see how many resources two PDs share for a given value of hops, e.g., what is the set of resources that are shared in the 3-hop radii of two PDs.
If a pair of PDs share fewer resources at a given number of hops than another pair of PDs, we can
say that the former pair is more isolated than the latter.

\subsection{Existing and New Mechanisms}
In \autoref{sec:model::model}, we showed with some examples that
\system is rich enough to capture existing mechanisms.
For example, unikernels are similar to virtual machines in many respects, but
the distinction between them is clearer in \system.
The application and kernel belong to the same PD, whose resources and
resource directory is a subset of the union of a conventional virtual machine
and process implementation.
Similarly, building slight variations of existing mechanisms is trivial.
For instance, to build processes that operate on a separate set of
physical pages, \system assigns different resource directory
entries (with a disjoint set of pages) to the PDs of those two processes.
Lightweight contexts (LwC)~\cite{litton2016light} are equally straight forward;
each LwC is a separate PD, but the various PDs share only the necessary
resources, e.g., virtual memory, files.

\subsection{Viewing Isolation as spectrum}
With \system, it is possible to provide different isolation levels for
different resources.
By varying $n_1, n_2$ and $\delta$ in \emph{NHopIsolated}
shown in~\autoref{sec:model::spectrum},
we show that there exists a
vast high-dimensional space of isolation primitives created by assigning
different isolation levels to different resources.
When deploying a new PD in a shared cloud environment, the operator
can vary these three parameters against other trusted and untrusted PDs.
And for a given deployment, the operator can use \emph{IsolationLevel} to determine
the degree of isolation between two untrusting PDs.
For example, if a new threat is discovered in the networking stack,
\system enables the deployment engineer to run just the
networking stack with an additional isolation level
until a vulnerability is patched.


\section{Conclusion}
\label{sec:conclusion}

We identify the problem that there is a lack of understanding about the level of
isolation and sharing provided by different isolation mechanisms.
Additionally, the plethora of isolation mechanisms do not share an underlying
framework, which makes it challenging to build new mechanisms.

We present the \system model, which lets us reason about isolation between applications
in a principled way.
It lets us model precisely which parts of the known and unknown software
state are shared between applications.
We then present the \system framework, designed to
realize the model by identifying the essential building blocks.

Finally, we show how \system lets us model and build new and existing
mechanisms that are precisely tailored to the user's isolation requirements,
and view the isolation of resources as a
spectrum.

\begin{acks}

We would like to thank Thomas Pasquier, Sam Leffler, and
George Neville-Neil for providing feedback on our earlier draft.

\end{acks}


\bibliographystyle{ACM-Reference-Format}
\bibliography{content/references}

\ifisdraft
Columnwidth: \the\columnwidth
\fi

\end{document}